\documentclass[amsfonts,showpacs,tightenlines,aps,12pt,floatfix]{revtex4}
\usepackage{bm}
\usepackage{epsfig}
\usepackage{amssymb,bm,mathrsfs,bbm,amscd}
\usepackage[tbtags]{amsmath}

\begin{document}


\title{Status of the Lattice and $\tau$ Decay Determinations of $\alpha_s$ 
\thanks{Work supported by Natural Sciences and Engineering Council of Canada}}

\author{K. Maltman}\email[]{kmaltman@yorku.ca}%

\affiliation{Department of Mathematics and Statistics, York University,
4700 Keele St., Toronto, ON CANADA M3J 1P3}
\altaffiliation{CSSM, Univ. of Adelaide, Adelaide, SA 5005 AUSTRALIA}

\begin{abstract}
The two highest precision determinations of $\alpha_s(M_Z^2)$,
that based on the analysis of short-distance-sensitive lattice
observables, and that based on an analysis of hadronic $\tau$ decay
data, have, until very recently, given results which are not
in good agreement. I review new versions of these analyses
which bring the two determinations into excellent agreement, and discuss
prospects for additional future improvements.
\end{abstract}


\maketitle




\section{Introduction}
Until recently, the determination of
$\alpha_s(M_Z^2)$ from perturbative analyses of short-distance-sensitive
lattice observables (yielding $0.1170(12)$~\cite{latticealphas05}), and
that from finite energy sum rule (FESR) analyses of
hadronic $\tau$ decay data (yielding $0.1212(11)$~\cite{davieretal08}),
produced central values which, though nominally precise, differed
from one another by $\sim 3\sigma$. In the past year, this discrepancy 
has been removed by new versions of both analyses.
In what follows, I outline the important features of these updates
which are responsible for this change.

\section{The Lattice Determination}
The original lattice determination~\cite{latticealphas05} involved
the perturbative analysis of various lattice observables, $O_k$,
computed using the MILC $n_f=2+1$ configurations. The $D=0$ expansions,
computed to 3-loops 
for the MILC action by Mason {\it et al.}~\cite{latticealphas05,hpqcd08},
have the form
\begin{equation}
O_k=\sum_{N=1}{\overline{c}}_N^{(k)}\alpha^N_V(Q_k)
\equiv D_k\alpha_V(Q_k)\sum_{M=0}c_M^{(k)}\alpha^M_V(Q_k)
\label{3loopPT}\end{equation}
where $\alpha_V$ is defined in the recent update~\cite{hpqcd08} (HPQCD08), 
and $Q_k=d_k/a$ is the relevant BLM scale. The ${\overline{c}}_{1,2,3}^{(k)}$
(equivalently, $D_k, c_1^{(k)}$, $c_2^{(k)}$), and $d_k$
are tabulated in HPQCD08. Regarding possible higher $D$ 
contributions, (i) $m_q$-dependent contributions were
removed by extrapolation, using data, and (ii) non-perturbative (NP) 
$m_q$-independent higher $D$ contributions were initially assumed to be
dominated by $D=4$ gluon condensate terms, which were then fitted and removed
independently for each $O_k$. For the lattice spacings, 
$a\sim 0.18,\, 0.12$, and $0.09\ fm$, of the original analysis, 
the observed scale-dependence of the $O_k$ could be reproduced only
by also fitting at least one additional coefficient
in Eq.~(\ref{3loopPT})~\cite{latticealphas05}. 

The updated HPQCD08~\cite{hpqcd08} and CSSM~\cite{mlms08} analyses incorporate
data from new MILC ensembles with $a\sim 0.15$ and $0.06\ fm$, with 
one very new $a\sim 0.045\ fm$ ensemble also employed in HPQCD08.
Useful cross-checks are also provided by
differences in (i) the choice of coupling employed and (ii) the treatment
of $m_q$-independent NP contribution in the two re-analyses. 
The choice of coupling in HPQCD08
leaves residual perturbative uncertainties in the conversion
from the $V$ to ${\overline{MS}}$ scheme, that in CSSM
in the effects of the truncated $\beta$ function, which
can be suppressed by focussing on finer lattices~\cite{mlms08}.
HPQCD08 performs an improved treatment of $m_q$-independent 
NP contributions, fitting a range of $D\geq 4$ forms to data,
while CSSM restricts its attention
to observables where the corresponding $D=4$ contributions,
estimated using charmonium sum-rule input for
$\langle \alpha_s G^2\rangle$~\cite{newgcond4}, are found to be
small. Even with the finer lattice scales of the new analyses, 
it turns out that at least one additional coefficient in
Eq.~(\ref{3loopPT}) must be fit. The $\alpha_s$ which results
provides an excellent representation of the scale
dependence of all the $O_k$ employed. The results of the two
re-analyses are in good agreement, and differ by only $\sim 1\sigma$
from the results of the earlier lattice analysis.
The equivalent $n_f=5$ results, at the scale $\mu^2 = M_Z^2$, are shown 
in Table~\ref{tablelattice}
for the three most- and four least-perturbative of the
$O_k$ studied in HPQCD08. $W_{kl}$
is the $k\times l$ Wilson loop and $u_0=W_{11}^{1/4}$.
Also shown is the corresponding quantity $\delta_{D=4}$, equal to
the percent shift in the scale dependence between
$a\sim 0.12$ and $a\sim 0.06\ fm$ resulting from first computing
$O_k$ using raw simulation values and then recomputing it after 
subtracting the known leading order $m_q$-independent $D=4$ 
contributions, estimated using charmonium sum rule input for 
$\langle \alpha_s G^2\rangle$~\cite{newgcond4}. $\delta_{D=4}$
provides a measure of the expected importance of the $m_q$-independent NP 
subtractions relative to the $D=0$ contribution from which $\alpha_s$ is 
to be determined. We see that residual NP effects in $O_k=log(W_{11})$ are 
expected to be tiny; the resulting subtraction in fact produces a shift of 
only $0.0001$ in $\alpha_s(M_Z^2)$~\cite{mlms08}. The fact that the results
for $\alpha_s$ obtained after fitting and subtracting what are expected
to be rather sizeable NP contributions to the most non-perturbative of 
the $O_k$ agree so well with those obtained from analyses of the data
for those $O_k$ where these corrections are expected to be very small
gives increased confidence in the treatment of such NP contributions
and, even more important, enhanced confidence in the reliability of
the results obtained from the most UV-sensitive of the $O_k$ 
considered.

\begin{center}
\begin{table}
\caption{ \label{tablelattice}  $\alpha_s(M_Z^2)$ and $\delta_{D=4}$
for the 3 least- and 4 most-non-perturbative of the $O_k$.}
\footnotesize
\begin{tabular}{|lllc|}
\hline 
$\qquad O_k$&\ \ $\alpha_s(M_Z^2)$&\ \ $\alpha_s(M_Z^2)$&$\delta_{D=4}$\\
&\ HPQCD08&\ \ CSSM&\\
\hline
$\log\left( W_{11}\right)$&$0.1185(8)\ \quad$&$0.1190(11)$\quad&$0.7\%$\\
$\log\left( W_{12}\right)$&$0.1185(8)\ \quad$&$0.1191(11)$\quad&$2.0\%$\\
$\log\left( {\dfrac{W_{12}}{u_0^6}}\right)$
&$0.1183(7)\ \quad$&$0.1191(11)$\quad
&$5.2\%$\\
\hline
$log\left({\dfrac{W_{11}W_{22}}{W_{12}^2}}\right)$
&$0.1185(9)\ \quad$&$\ \ \ N/A$\quad&$32\%$\\
$log\left({\dfrac{W_{23}}{u_0^{10}}}\right)$
&$0.1176(9)\ \quad$&$\ \ \ N/A$\quad&$53\%$\\
$log\left({\dfrac{W_{14}}{W_{23}}}\right)$&$0.1171(11)\quad$&$\ \ \ N/A$\quad
&$79\%$\\
$log\left({\dfrac{W_{11}W_{23}}{W_{12}W_{13}}}\right)$
&$0.1174(9)\ \quad$&$\ \ \ N/A$\quad&$92\%$\\
\hline
\end{tabular}
\end{table}
\end{center}

\section{The hadronic $\tau$ decay determination}

In the SM, with $\Gamma^{had}_{V/A;ud}$ the $I=1$ V or A current-induced 
width for $\tau$ to hadrons, $\Gamma_e$ the $\tau$ electronic width,
$y_\tau =s/m_\tau^2$, $S_{EW}$ a known short-distance EW correction,
and $R_{V/A;ud}=\Gamma^{had}_{V/A;ud}/\Gamma_e$, one has~\cite{tsai}
\begin{eqnarray}
dR_{V/A;ud}/dy_\tau &=&12\pi^2S_{EW} \vert V_{ud}\vert^2
\left[ w_{00}(y_\tau )\rho^{(0+1)}_{V/A}(s)\right.\nonumber\\
&&\left.\ \ -w_L(y_\tau )\rho_{V/A}^{(0)}(s)
\right] ,
\label{taubasic}\end{eqnarray}
where $\rho^{(J)}_{V/A}(s)$ are the spectral functions of the
spin $J$ scalar correlators, $\Pi_{V/A}^{(J)}(s)$,
of the $I=1$, V/A current-current 2-point functions, 
$w_{00}(y)=(1-y)^2(1+2y)$, $w_L(y)=2y(1-y)^2$ and,
up to $O\left[ (m_d\pm m_u)^2\right]$ corrections,
$\rho^{(0)}_{V}(s)=0$ and $\rho_{A}^{(0)}(s)=2f_\pi^2\delta
(s-m_\pi^2)$,
making $\rho^{(0+1)}_{V/A}(s)$ accessible from the
experimental distributions $dR_{V/A;ud}/dy_\tau$~\cite{alephud05,opalud99}.
For any $s_0$ and any analytic $w(s)$, the related finite energy
sum rule (FESR)
\begin{equation}
\int_0^{s_0}w(s) \rho^{(0+1)}_{V/A}(s) ds\, =\, -{\frac{1}{2\pi i}}
\oint_{\vert s\vert =s_0}w(s) \Pi^{(0+1)}_{V/A} (s) ds,
\label{basicfesr}
\end{equation}
is satisfied. For large enough $s_0$, the OPE can be employed on the RHS.
For typical $w(s)$, and $s_0$ above $\sim 2\ {\rm GeV}^2$,
the OPE is strongly $D=0$ dominated and hence largely determined by 
$\alpha_s$. The 5-loop version of the $D=0$ OPE series~\cite{bck08} 
is employed in all recent $\tau$-based $\alpha_s$
analyses~\cite{bck08,davieretal08,cgp08,jb08,kmty08,narison09,menke09}.
Use of polynomial weights, $w(y)$, with $y=s/s_0$, helps to quantify
higher $D$ contributions, most of which must be fit to data, since (i) up 
to corrections of $O\left(\alpha_s^2\right)$, the integrated OPE series
terminates at $D=2N+2$ (with $N$ the degree of $w(y)$), and 
(ii) integrated OPE contributions of different $D$ scale differently
with $s_0$ ($D=2k+2$ terms scaling as $1/s_0^k$).

$\tau$ decay determinations of $\alpha_s$ have conventionally been based 
on a combined analysis of the $s_0=m_\tau^2$, $km=00$, 
$10$, $11$, $12$, $13$, $w_{km}(y)=w_{00}(y)\, (1-y)^ky^m$ ``spectral 
weight FESRs''~\cite{alephud05,davieretal08,opalud99}. This analysis 
relies crucially on the assumption that 
$D=10,\cdots ,16$ contributions, each in principle present for one or 
more of the $w_{km}$ employed, can, in all cases, be safely neglected. 
A number of recent analyses employ either this strategy
directly~\cite{davieretal08} or make use of nominal $D=6,8$ NP
contributions extracted by doing so~\cite{bck08,jb08,cgp08,menke09}.
This assumption is, however, potentially dangerous since (i) a 
$\sim 1\%$ determination of $\alpha_s(M_Z^2)$ requires control of $D>4$
 NP contributions to $\lesssim 0.5\%$ of the leading $D=0$ term and 
(ii) the $km=11$, $12$ and $13$ FESRs have strongly suppressed $D=0$ 
OPE contributions. The validity of the assumption that all $D>8$
contributions can be neglected was tested in MY08~\cite{kmty08} by
(i) studying, as a function of $s_0$, the match between
the variously weighted 
OPE integrals, evaluated using fitted OPE parameters, and the corresponding 
experimental spectral integrals, and (ii) using the same data and
fitted OPE parameters as input to FESRs for different $w(y)$ involving the 
same set of OPE parameters. The quality of the match produced by
the results of the optimized $w_{km}$ analysis of the ALEPH data
was found to be typically rather poor in the window 
$\sim 2\ {\rm GeV}^2<s_0\leq m_\tau^2$, not just for the $w_{km}$ 
employed in the analysis, but also for other degree $\leq 3$ weights, 
whose OPE integrals 
should depend only on the $D=0,4,6,8$ OPE parameters included in 
the ALEPH fit~\cite{kmty08}. Similar discrepancies exist for the
$w_{km}$ analysis based on the OPAL data, and for the BJ08~\cite{jb08} 
treatment of the $w_{00}$ FESR, which employs a different set of assumed 
values for the $D=6,8$ contributions~\cite{kmty08}.

In view of these problems, MY08 performed analyses based
on alternate weights, $w_N(y)=1-{\frac{N}{N-1}}y+{\frac{1}{N-1}}y^N$,
designed to suppress $D>4$ relative to the leading $D=0$ 
contributions and hence optimize the determination of $\alpha_s$.
(In terms of its size relative to the leading, $\alpha_s$-dependent integrated
$D=0$ series, neglect of $D>8$ contributions would, in fact, be between $7$ 
and $814$ times safer for the $w_N$ analyses than it is for the higher 
$w_{km}$ FESRs of the conventional analyses~\cite{kmty08}. The fact that
only a single $D>4$ contribution, with $D=2N+2$, enters the $w_N$
FESR also simplifies fitting the corresponding OPE contribution.)
One finds (i) excellent consistency among
the $\alpha_s$ values obtained using different $w_N(y)$ and/or
different channels (V, A or V+A); (ii) (as intended) a significantly
reduced impact of $D>4$ OPE contributions; and (iii) in contrast 
to the results of the combined $\{ w_{km}\}$ analysis, an excellent
quality match between OPE and corresponding spectral integrals
for other degree $\leq 3$ weights (including the kinematic weight
$w_{00}$) over the whole of the $s_0$ window noted above.
The results of MY08 are the only ones in the literature to
satisfy this set of self-consistency constraints.
Since (i) the $\{ w_{km}\}$ analyses, which should produce results in agreement
with those of the corresponding $w_N$ analyses when using the
same data, instead produce significantly larger $\alpha_s$, and
(ii) (as shown in the Figures of MY08) the results of the $\{ w_{km}\}$
analyses, considered at lower $s_0$, produce optimized OPE integrals not 
in agreement within errors with the corresponding experimental spectral 
integrals, and, moreover, significantly inferior to the matches obtained 
using the $\{ w_N\}$ analysis fit parameters, we take the results for 
$\alpha_s$ to be those obtained from the $\{ w_N\}$ analysis of MY08. The 
central value, $\alpha_s(m_\tau^2)=0.321(5)(12)$ (where the first error is 
experimental and the second theoretical) is based on the CIPT treatment
of the $D=0$ series, which yields better consistency among the results 
of the different $w_N$ FESRs than does the corresponding FOPT
treatment~\cite{kmtau08}. This corresponds to
\begin{equation}
\alpha_s(M_Z^2)=0.1187(6)(15)(3)\ ,
\end{equation}
where the errors are, respectively, experimental, theoretical, and
that due to evolution. The $\tau$ result
is now in excellent agreement with the lattice determination.
Note that effects associated with the breakdown of the integrated
OPE representation, if any, are not reflected in the errors quoted
here. There are, in fact, no signs for the presence of
such effects in the match between the optimized OPE and spectral
integrals at present. Further discussion of this issue
may be found in the next section.

\section{Discussion and prospects}
Given the sizeable estimated $D=4$ NP subtractions for the most
non-perturbative of the lattice observables $O_k$, and the necessity
of fitting analogous $D>4$ contributions to data in an approximate way,
the results corresponding to those $O_k$ expected to have very
small $m_q$-independent NP contributions should be
the most reliable sources of $\alpha_s$ in the lattice analysis.
The difference between the result obtained from the most UV-sensitive of 
the observables, $O_k=log(W_{11})$, $\alpha_s(M_Z^2)=0.1185(8)$, and 
that representing the average over the results for all the
$O_k$ considered in HPQCD08, $\alpha_s(M_Z^2)=0.1183(8)$, 
is however, in fact, very small on the scale
of the uncertainties of the analysis. We note also that
one would have to reduce the lattice spacing by roughly an order of magnitude
before perturbative coefficients beyond 3-loops, which must be fitted at 
present, could be plausibly neglected. This is certainly
not feasible. Fortunately, the fitting
procedures employed, when applied to input pseudo-data generated using
a known high-order input expansion, return the input $\alpha_s$
value with very good accuracy. As a result, it appears highly
unlikely that the errors associated with the fitting of the
higher order coefficients play
any significant role. The uncertainty in $\alpha_s$ for what we
would argue is the most favorable of the $O_k$ analyses,
that based on $O_k=log(W_{11})$, as can be seen from
the results quoted by HPQCD08, will thus be dominated by the uncertainties
in the determination of the lattice spacings in physical
units for the various ensembles, with comparable contributions coming from
the uncertainties in $r_1/a$ and $r_1$ itself~\cite{hpqcd08}. It
appears unlikely that these can be significantly reduced in the
near future, so one should expect only minor improvements in
the lattice determination, such as those that should result from access to
a larger set of $a\sim 0.045$ fm ensembles. 

The current errors on the $\tau$ determination are larger than for
the lattice determination, but may
be amenable to more significant near-term reduction. The major
components of the quoted $0.012$ uncertainty on $\alpha_s(m_\tau^2)$
are: $0.0084$ from the maximum difference between CIPT and FOPT
determinations over the $w_2,\cdots ,w_6$ set considered in MY08;
$0.0059$ from the uncertainty in the $0.009(7)\ {\rm GeV}^4$ charmonium sum
rule input employed for $\langle \alpha_sG^2/\pi\rangle$; and
$0.0056$ from the assumed $100\%$ uncertainty on the FAC estimate
for the $6$-loop coefficient of the $D=0$ Adler function~\cite{bck08}.
We discuss these contributions briefly in what follows.

BJ08 have raised an interesting question about the relative reliability of the
FOPT and CIPT prescriptions for evaluating the truncated $D=0$ series.
Based on a model reflecting known general properties
of the divergent $D=0$ series, they argue that, despite better observed
convergence behavior for the truncated CIPT series, the truncated FOPT
series might, nonetheless, better approximate the true result. 
Currently, it is known that, contrary to what one would expect
from this model, a combined fit to a collection of degree $\leq 3$
weights using the truncated fifth order CIPT prescription yields
a good combined OPE-spectral integral match, 
while the analogous combined fit using the
truncated FOPT prescription does not~\cite{kmtau08}. The analogous
calculation, however, has yet to be carried out for the fully resummed
BJ08 model. This work is in progress. An interesting
additional observation is that the $w_2$ case yields a much better agreement
between the CIPT and FOPT treatments of the integrated $D=0$ OPE
series than is found for other weights, and that
this agreement persists over the whole of the $2\ {\rm GeV}^2<s_0<m_\tau^2$
window. A much better than usual agreement between the FOPT and CIPT versions
of the $w_2$-weighted integrated truncated $D=0$ OPE 
series in the vicinity of the smallest terms is also seen in the 
BJ08 model, suggesting that the $w_2$ FESR may be especially well suited
to the determination of $\alpha_s$. Further work, however, is needed
to have any realistic hope of significantly reducing the existing
FOPT vs. CIPT prescription-dependence of the results for $\alpha_s$. 
The prescription-dependence uncertainty is, of course, intimately 
related to that associated with the uncertainty on the estimated 
$6$-loop Adler function coefficient.

Reduction in the error associated with the uncertainty on the input
for $\langle \alpha_sG^2\rangle$ would require an improved determination of
this condensate. Given the associated renormalon ambiguity,
such a determination should be performed using the same hadronic 
$\tau$ decay data, in an analysis with a consistently truncated $D=0$ series. 
Preliminary studies indicate that it is, in fact, likely that such an 
improved determination of $\langle \alpha_sG^2\rangle$ is possible. 
This issue turns out to be closely linked to the question of the level 
of residual duality violation (integrated OPE breakdown)
present in the FESRs employed in the analysis, and the modelling
of duality violating contributions to the physical spectral functions. 

A final potential error source for the $\tau$ decay determination is
residual integrated duality violation.
While (i) it is known empirically
that, at the scales employed in the $\tau$ decay analyses, OPE
breakdown is localized to the vicinity of the timelike point on the
OPE contour~\cite{kmfesr98}, and (ii) the weights employed in the
analysis described above all have a double zero at the timelike
point, a physically well-motivated model for such
duality violations~\cite{cgp08}, when fitted
to the observed V and A spectral functions, allows duality-violating-induced
shifts to $\alpha_s(m_\tau^2)$ in the range $0.003-0.010$. An additional
uncertainty at the upper end of this range would  not be negligible on 
the scale of the other errors
quoted above. Preliminary investigations, however, indicate that significant
further constraints can be placed on this model and that, when they
are, (i) the allowed duality-violating-induced shifts lie at the low end
of the range quoted above, and (ii) an improved simultaneous determination 
of the gluon condensate is almost certainly possible. Further work is required 
before more quantitive statements on these issues can be made.

\end{document}